\iffalse
\documentclass[twoside,leqno,twocolumn]{article}

\usepackage[letterpaper]{geometry}
\usepackage{ltexpprt}
\usepackage{stix}
\else
\documentclass[leqno,twocolumn]{article}

\usepackage[a4paper]{geometry}
\newtheorem{theorem}{Theorem}
\usepackage{stix}
\advance\textheight by 3cm
\advance\topmargin -1.5cm

\advance\textwidth by 3cm
\advance\oddsidemargin by -1.5cm
\advance\evensidemargin by -1.5cm
\fi

\usepackage{tikz,pgfplots}

\usepackage{pdfsync}
\usepackage{amsmath, amssymb}
\usepackage{graphicx}
\usepackage[hyphens]{url}
\usepackage{mathrsfs}
\usepackage{bm}
\usepackage{todonotes}
\usepackage{booktabs}

\DeclareRobustCommand{\stirling}{\genfrac\{\}{0pt}{}}

\renewcommand\phi\varphi
\newcommand{\?}{\mskip1.5mu}
\usepackage{microtype} % micro-typographic extensions
\usepackage{algorithm}
\usepackage[noend]{algorithmic}
\newcommand{\Z}{\mathbf Z}
\newcommand{\lst}[2]{${#1}_0$,~${#1}_1$,$\dots\,$,~${#1}_{#2-1}$}

\usepackage{framed}

\usepackage[utf8]{inputenc}
\usepackage[T1]{fontenc}
\usepackage[english]{babel}
\usepackage{amsmath}
\usepackage{graphicx}
\usepackage{multirow}
\usepackage{verbatim}
\usepackage{amsfonts}
\usepackage{amssymb}
\usepackage{mathtools}
\usepackage{array}
\usepackage[page]{appendix}

\DeclarePairedDelimiter\floor{\lfloor}{\rfloor}
\DeclarePairedDelimiter\ceil{\lceil}{\rceil}

\def\..{\,\mathpunct{\ldotp\ldotp}} % Middle stuff for intervals. Usage: \..

\title{RecSplit: Minimal Perfect Hashing\\via Recursive Splitting}
\author{Emmanuel Esposito\thanks{Universit\`a degli Studi di Milano, Italy}\and Thomas Mueller Graf\thanks{Independent researcher, Switzerland} \and Sebastiano Vigna\thanks{Universit\`a degli Studi di Milano, Italy}}

\date{}
\begin{document}
\bibliographystyle{alpha}
\maketitle
\begin{abstract}\small\baselineskip=9pt
A \emph{minimal perfect hash function} bijectively maps a key set $S$ out of a universe $U$ into
the first $|S|$ natural numbers. Minimal perfect hash functions are used, for example, to map irregularly-shaped keys,
such as strings, in a compact space so that metadata can then be simply stored in an array.
While it is known that just $1.44$ bits per key are necessary to store a minimal perfect hash function,
no published technique can go below $2$ bits per key in practice.
We propose a new technique
for storing minimal perfect hash functions with expected linear construction time and expected
constant lookup time that makes it possible to build for the first time, for example, structures which need $1.56$
bits per key, that is, within $8.3$\% of the lower bound, in less than $2$\,ms per key.
We show that instances of our construction are able to \emph{simultaneously} beat the construction time,
space usage and lookup time of the state-of-the-art data structure reaching $2$ bits per key.
Moreover, we provide parameter choices giving structures
which are competitive with alternative, larger-size data structures in terms of space and lookup time.
The construction of our data structures can be easily
parallelized or mapped on distributed computational units (e.g., within the MapReduce framework),
and structures larger than the available RAM can be directly built in mass storage.
\end{abstract}

\section{Introduction}

\emph{Minimal perfect hash functions (MPHFs)} are static data structures
storing a bijection of a given set $S$ of keys, $|S|=n$, into
the first $n$ natural numbers.
While such a bijection can easily be stored using hash tables, MPHFs are allowed
to return \emph{any} value if the queried key is not in the original set $S$;
this relaxation enables to break the information-theoretical lower bound of
storing the set $S$.
Indeed, MPHFs constructions achieve $O(n)$ bits of space, regardless of the size
of the keys. This property makes MPHFs powerful techniques when handling, for
instance, large sets of strings, and they are important building blocks of
space-efficient data structures.

The usual requirement on MPHFs is that construction can be completed in linear time,
and that lookup takes constant time; often these requirements are relaxed to
\emph{expected} linear/constant time.

The space lower bound of a MPHF is about $\lg e\approx 1.44$
bits per key~\cite{FKSSST}, and independent from $S$. Some theoretical constructions
reach the lower bound, but they work only for preposterously large
$n$~\cite{HTEMPHNMS}. In practice, the CHD construction~\cite{BBDHDC}, detailed
below, is currently able to reach about $2$ bits per key, but at such a space we show that it
is significantly slower than our approach for both construction and lookup.
Techniques based on random linear
systems, such as derivatives of the venerable construction by Majewski, Wormald, Havas, 
and Czech~\cite{MWHFPHM}, are reasonably
fast in construction and quite fast in lookups, but they each only achieve
about $2.24$ bits per key~\cite{GOVFSCF}. Finally, techniques based on
fingerprints~\cite{MSSRPHUF} are extremely fast in construction and lookup,
but only when the space occupancy is extremely large (e.g., above $4$ bits per
key).

In this paper we present a new technique for storing minimal perfect hash
functions based on recursive splitting and brute-force searching, with expected
linear construction time and constant expected lookup time.
Our technique is able to break the $2$ bits barrier, and at the
same time can improve the construction time, space usage and lookup time of CHD. 

Our approach to build an MPHF is conceptually simple, and on a high level similar to existing approaches:
we partition a set into buckets, and then process each bucket independently.
Unlike other approaches, in our case the entries of a bucket logically form a tree which
defines an independent MPHF.
We analyze how exactly those trees should "look like",
for a brute-force search to be efficient, for lookup to be fast, and to use little space.
We also show how to implicitly encode such trees, and how to
employ succinct data structures that allow efficient lookup and storage 
even for very small MPHFs.

One of the main virtues of our approach is that it makes it possible
to create practical MPHFs using less than $9$\% more space than the information-theoretical
lower bound, compared to the state of the art, which uses $40$\% more space.
But our approach is also very flexible and amenable,
such that it is useful for a wide range of parameters:
it not only yields the most space-saving practical MPHFs,
but it is also competitive regarding construction time and lookup
over a wide range of parameter values.
Existing algorithms, on the other hand, are competitive only in a narrow area.

All the code used in this paper is available from the authors under the
GNU Lesser General Public License, version 3 or later, as a part of the
Sux project (\url{http://sux.di.unimi.it/}).

\section{Notation}

We use von Neumann's definition and notation for the set of natural numbers,
so $n=\{\?0,1,\ldots,n-1\?\}$.
We use $\gg$ and $\ll$ for left and right bit shifting.
Following Knuth~\cite{KnuACPIV}, we
use $\lambda x$ for the index of the highest
(leftmost) bit set ($\lambda x$ is undefined when $x\leq 0$).

\section{Background and related work}

\subsection{Early work} Sprugnoli~\cite{SprPHF} defines perfect hash functions
on small sets by testing exhaustively constants in simple expressions involving
integer divisions and remainders; in some small-sized cases, he is able to find
a MPHF. He suggests that for large key sets one can first hash the keys into a
fixed number of buckets (called therein \emph{segments}), and then operate
independently. Jaeschke~\cite{JaeRH} refines his approach to obtain minimal perfect hash
functions on small sets by exhaustive search.
Both exhaustive search and hashing into buckets are essential elements of our construction.

\subsection{Random Linear Systems}
\label{sec:MWHC}

In their seminal paper~\cite{MWHFPHM}, Majewski, Wormald, Havas, and Czech (MWHC
hereinafter)
introduced the first compact construction for static functions
using the connection between linear systems and hypergraphs. To store a
function $f:S\to t$, they generate a random system with
$n=|S|$ equations in $m$ variables of the form
\begin{equation*}
%\label{eq:MWHC}
 w_{h_0(x)}+ w_{h_1(x)}+ \cdots
+ w_{h_{k-1}(x)}= f(x)\mod t \quad x \in S.
\end{equation*}
Here $h_i:U\to m$ are $k$ fully random hash
functions, and the $w_i$'s are variables assuming values in $\Z/t\Z$.
Due to bounds on the acyclicity of random graphs, if the ratio between the
number of variables and the number of equations is above a certain threshold
$\gamma_k$, the system can be almost always
triangulated in linear time by peeling the
associated hypergraph, and thus solved; in
other words, we have both a probabilistic guarantee that the system is
solvable, and that the solution can be found in linear time.

The data structure is then a solution of the system (i.e., the values of the
variables $w_i$): storing the solution makes it possible to compute
$f(x)$ in constant time. The space usage is approximately $\gamma_k b$ bits per key.
The constant $\gamma_k$ depends on the degree $k$, and attains its minimum at
$k=3$ ($\gamma_3\approx 1.23$).

Chazelle, Kilian,
Rubinfeld and Tal~\cite{CKRBF}, unaware of the MWHC construction,
proposed it independently, but also exploited the fact that a peelable hypergraph
is also orientable: since as a side effect of the peeling
process each hyperedge
can be assigned a unique vertex (or, equivalently, each
equation can be assigned a unique variable), each key can be assigned injectively
an integer in $\lceil\gamma_k n\rceil$.
Then, we just need to modify the MWHC construction so that instead of storing $f(x)$, we store
which of the $k$ vertices of the hyperedge is the assigned one, and this can be
done in approximately $\gamma_k\lceil \log k \rceil$ bits per key. At
retrieval time, the value we store makes it possible to recover the unique integer
assigned to the key.

To make the perfect hash function minimal, that is, a
function to $n$, rather than $\lceil\gamma_r n\rceil$, a
\emph{ranking} data structure can be added. A bit vector of size $\lceil\gamma_r n\rceil$
records which vertices have been assigned to a hyperedge, and then with
additional $o(n)$ bits the number of ones preceding a given one
(i.e., its rank) can be computed in constant time (see~\cite{VigBIRSQ} for practical implementations).

Again, the best $k$ is $3$, which yields theoretically a $2.46n + o(n)$ bits data
structure~\cite{BPZPPHNOS}, using $2$ bits per variable: since
the value $3$ never appears in the solution, it
can be used instead of zero to mark vertices associated with hyperedges. In this
way, vertices assigned to hyperedges have always nonzero values and are the
only ones with nonzero values because of the way the solution is computed
by the peeling process. In the end, it is
easy to adapt ranking structures so as to rank nonzero pairs of bits,
rather than ones, eliminating the need for the additional bit vector (see~\cite{BPZPPHNOS} for details).
Finally, Genuzio, Ottaviano and Vigna have shown that it is possible
to use bounds on \emph{solvability}, rather than \emph{acyclicity}, for the
same purpose, obtaining a $2.24n + o(n)$ bits data
structure~\cite{GOVFSCF}.

\subsection{CHD}\label{sec:CHD} Belazzougui, Botelho, and Dietzfelbinger~\cite{BBDHDC} introduced a completely
different construction, called CHD (compressed hash-and-displace), which
makes it possible, in theory, to reach the
information-theoretical lower bound of $\approx 1.44$ bits per key,
at the price of increasing the expected construction time. Keys are first
mapped into small buckets of expected size $\lambda$ (e.g., $\lambda=4$), and then buckets are mapped into
the codomain without collisions, starting with the largest ones. To map the buckets
into the codomain they examine for each bucket an enumeration of fully random independent
hash functions $\phi^k_i:U\to k$, $i=0,1,2,\ldots$: once a function mapping
the current bucket in the codomain without collisions is found, its index
is stored in the data structure. While the theoretical analysis suggests
that by increasing the bucket size it is possible to achieve space usage
as close to the lower bound as desired, in practice it is unfeasible to go below
$2$ bits per key.

\subsection{Fingerprinting}\label{sec:finger} Recently, M\"uller~\textit{et al.}~\cite{MSSRPHUF}
introduced a completely new technique for minimal perfect hashing. A series of
bit arrays of decreasing size, called \emph{levels}, is used to record
information about collisions between keys. More precisely, all positions in the
first level to which a single key is mapped by a random hash function
are marked with a one. Then, one
takes all keys which participated in collisions, and tries to map them into the
second level, using the same strategy, and so on. As a result, one obtains a perfect
hashing of the key set: to retrieve the output associated with a key, one
maps the key in turn to each level until a one is found, and then the (overall)
position of the one yields a distinct number for each key. A constant-time
ranking structure over the concatenation of the bit arrays is then used, as in
the case of hypergraph-based techniques, to turn the perfect hash function into
a minimal perfect hash function.

Fingerprint-based minimal perfect hash functions have the advantage of being
very tunable for speed (both for construction and lookup)
by suitably setting the length of each level, and thus the space usage: at $3$
bits per elements, for example, the authors report that only $1.56$ levels are accessed
on average, resulting in a very low number of cache misses. However, the
best result for space is $2.58$ bits per key, which is not competitive,
for example, with CHD or the techniques described in this paper. Moreover,
at that size construction and lookup are very slow.

% \section{Statement of the problem and previous work}
%
% Given a set of $n$ keys $S$ out of a universe $U$, the \emph{minimal perfect hash function}
% problem is that of storing a bijective function $f:S\to n$ so that given $x\in U$ we can compute
% in (expected) constant time a value that will be exactly $f(x)$ when $x\in S$. Construction
% should happen in (expected) linear time.
%
% The key point here is that we give no condition when $x\not\in S$ (except that
% the computation must still be performed in constant time), so the
% information-theoretical lower bound for storing $S$ as a subset of $U$ does not
% apply.
% The function is not specified, so the lower bound per key is very small
% (about $\approx 1.44$ bits~\cite{FKSSST}),
% and independent from $S$. Some theoretical constructions reach the lower
% bound, but they work only for preposterously large $n$~\cite{HTEMPHNMS}.

\section{RecSplit}

RecSplit is based on the idea that for very small
sets it is possible to find a MPHF simply by brute-force searching for
a \emph{bijection} with suitable codomain. However, we
extend the idea to brute-force searching of \emph{splittings} which bring large sets
to an amenable size.

To build a RecSplit instance for a set of $n$ keys, we first distribute keys randomly into buckets
of average size $b$ using a random hash function $g:S\to \lceil n/b\rceil$. Then,
each bucket is recursively \emph{split} into smaller pieces until we reach
a target \emph{leaf size} $\ell$ on which it is feasible to directly search
for an MPHF. As we will see, the parameters $\ell$ and $b$ provide different
space and time tradeoffs. We will build an MPHF
for each bucket independently, opening the way to parallel or distributed
construction.

\begin{figure*}[t]
\centering
\includegraphics[scale=1]{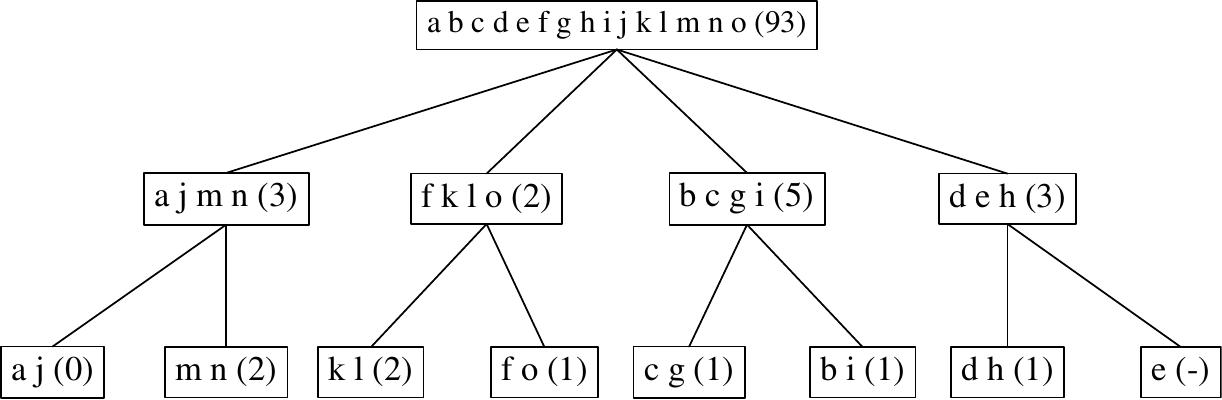}
\caption{\label{fig:tree}An example of a splitting tree for $15$ element with $\ell=2$. Each
node is decorated with the associated keys and the splitting or bijection index. Note that the leaf of size one has no associated index.}
\end{figure*}

We search for both splittings and bijections considering an enumeration of fully random independent
hash functions $\phi^k_i:U\to k$, $i=0,1,2,\ldots$, whose indices will be stored
in the data structure.

Given a set of keys $X$ (e.g., a whole bucket) of size $m$,
a split is defined by a list of parts \lst ks such that $\sum_{i=0}^{s-1}k_i=m$.
We compute the splitting index by searching for the first function $\phi^m_i$
such that
\[\Biggl|\bigl(\phi^m_i\bigr)^{-1}\Biggl(\,\biggl[\sum_{i=0}^{t-1}k_i\..
\sum_{i=0}^tk_i\biggr)\,\Biggr) \cap X\Biggr| = k_t,\quad 0\leq t < s-1.\]
In other words, we map the elements of
$X$ in $m$, and find the first function such that the elements mapped to a value smaller
than $k_0$ are exactly $k_0$, the elements mapped to values greater than or
equal to $k_0$ but smaller than $k_0+k_1$ are exactly $k_1$, and so
on.

Then, we proceed recursively on the $s$ parts until the current size
of a part is less than or equal to $\ell$.
At that point, given a part $X$ we search for the first function $\phi^{|X|}_i$ that is bijective on $X$.
Thus, we obtain a rooted \emph{splitting tree} with a \emph{splitting index} associated to each
internal node and a \emph{bijection index} associated with each leaf (see Figure~\ref{fig:tree}). We will represent
each such index using an optimal Golomb-Rice instantaneous code~\cite{SolDC}, as such indices have
a geometric distribution.

While the size of each part is in principle arbitrary, we will see that it is convenient
to work with sizes that are multiples of $\ell$, and to have all
parts of the same size except possibly the last one.
In this way, in every bucket \emph{all leaves have
exactly size $\ell$}, except possibly the last one, and for each tree node \emph{all subtrees have
the same shape}, except possibly the last one.

A \emph{splitting strategy} is made of a leaf size $\ell$ and a desired
\emph{splitting unit} $u$ (which must be a multiple of $\ell$) for each $m>\ell$. These pieces of information univocally define
the shape of the splitting tree for each $m>\ell$: the fanout of a node
associated with $m$ keys will be $\lceil m/u\rceil$, with all parts of size $u$ except possibly
the last one, which may have size $m\bmod u$.

Given a splitting tree for a set $X$, we can compute a MPHF on $X$:
given $x\in X$, we follow the tree from the root up to a leaf using the splitting
indices. Every time we move to a child, we know exactly how many keys we are leaving to
our left, as that number only depends on the splitting strategy. Thus, when we
get to a leaf, we know that, say, $c$ keys are mapped to the left of the
leaf. We then apply the bijection associated with the leaf to $x$, and add $c$ to the
resulting value. Multiple buckets are handled by keeping track of the prefix sums of the number of
keys in each bucket (i.e., of how many keys are assigned to previous buckets) and modifying
the result accordingly.

\section{Searching for splittings and bijections}

Many different splitting strategies are possible. While a completely
analytical analysis appears to be formidable, in this section we prove a number
of results which will make it much easier to devise a good strategy.

A first observation is that bucket sizes are distributed as a
binomially distributed random variable $S$ with
parameters $p=b/n$ and $n$. The moments of $S$ can be written as~\cite{KnoCFEMBPD}
\begin{multline*}
\mathbf E\bigl[S^d\bigr] = \sum_{i=0}^d \stirling di i! {n\choose i}\Biggl(\frac bn\Biggr)^i\\=
\sum_{i=0}^d \stirling di \frac{n!}{n^i(n-i)!} b^i\leq \sum_{i=0}^d \stirling di  b^i.
\end{multline*}
In particular, every algorithm that is polynomially bounded (even just in expectation) for each bucket will be linear
in expectation when run on all keys, and any lookup algorithm doing polynomial work on a bucket
will be constant-time in expectation.
% \footnote{These statements hide the fact that for any practical
% $n$ and $b$ the expected number of buckets larger than, say, $6b$ is much smaller than one.}

\subsection{Searching for splittings}

Assuming that our family of functions is fully random, the probability of
finding a split for a set of size $m$ in a left part of size $k$ and a right part of size $m-k$ is
\begin{multline*}
\frac1{m^m}{m\choose k} k^k(m-k)^{m-k}\sim\\
\frac1{m^m}\frac{\bigl(\frac me\bigr)^m\sqrt{2\pi
m}}{\bigl(\frac{k}e\bigr)^k\sqrt{2\pi k} \bigl(\frac{m -
k}e\bigr)^{m-k}\sqrt{2\pi (m-k)}} k^k(m-k)^{m-k}\\=\sqrt{\frac{m}{2 \pi k(m-k)}}.
\end{multline*}
using Stirling's approximation.
Then, the average number of trials to find a splitting hash function will be
asymptotic to
\[
\sqrt{\frac{2\pi k(m-k)}m}
\]
which is maximized when $k = \lceil m / 2\rceil$, that is,
for balanced splits.
% The number of function evaluations will be bounded by
% \[
% m \sqrt{\frac{2\pi k(m-k)}m} =\sqrt{2\pi m k(m-k)}.
% \]
% This bound is quite rough, as usually one realizes that the split
% is unbalanced before all elements have been tested.

We remark that our choice of codomain $m$ and threshold $k$ is optimal. Indeed,
if we consider the general case of a codomain $r$ and a threshold $t$, the
probability
\begin{equation}
\label{eq:rt}
\frac1{r^m}{m\choose k} t^k(r-t)^{m-k}
\end{equation}
depends only on the ratio $r/t$, as
\[
\frac1{(\alpha  r)^m}{m\choose k} { (\alpha  t)}^k( \alpha  r-
\alpha t)^{m-k} =\frac1{r^m}{m\choose k} t^k(r-t)^{m-k}.
\]
Now, if we explicitly write $t = \alpha r$ and minimize~(\ref{eq:rt}) in
$\alpha$, we obtain $\alpha=k/m$. Thus, any choice of $r$ and $t$ with $t/r=k/m$
provides the highest probability of success.

We will also use splittings with more than two parts. In general, if we want to
split $m$ into $s$ parts as $m = k_0+k_1+\cdots+k_{s-1}$, the number of splitting functions
is given by the number of possible ordered partitions of $m$ into sets of 
cardinality \lst ks (i.e., the associated multinomial coefficient), 
multiplied by the number of possible functions within each set. Dividing
by the overall number of functions $m^m$ we obtain
\begin{multline}
\label{eq:probsplit}
\frac1{m^m}{m \choose k_0,k_1,\ldots, k_{s-1}}\prod_{i=0}^{s-1}\bigl(k_i\bigr)^{k_i}
\\=\frac1{m^m}\frac{m!}{\prod_{i=0}^{s-1}k_i!}\prod_{i=0}^{s-1}\bigl(k_i\bigr)^{k_i}
	=\frac{m!}{m^m}\prod_{i=0}^{s-1}\frac{\bigl(k_i\bigr)^{k_i}}{k_i!}\\\sim
\sqrt{\frac m{(2\pi)^{s-1}\prod_{i=0}^{s-1} k_i}},
\end{multline}
where the last approximation is once again
obtained using Stirling's. In the end, the average number of trials is asymptotically
\begin{equation}
\label{eq:avgsplit}
\sqrt{\frac{(2\pi)^{s-1}\prod_{i=0}^{s-1} k_i}m}.
\end{equation}
By Jensen's inequality the worst case happens when all parts are equal, that is,
 \[
 \sqrt{\frac {(2\pi)^{s-1}\bigl(\frac ms\bigr)^s}m},
\]
which grows polynomially in $m$ for fixed $s$, but with exponent
$s$, which means that we cannot expect to be able to choose a large value
of $s$.

We now want to give a bound of the expected amount of work done on a single bucket.
To do so, we make a simplifying assumption, that is, that splitting always happens in $s$ parts:
we fix the leaf size $\ell$
and we use a bucket size of $s^h \ell$ (if the bucket has not a size of this
form, we use the closest approximation from above). Let $X^i_j$ be a set of random variables,
where $i$ represent the distance from the root, and $0\leq j\leq s^i-1$. The variables
are independent and for each $i$ they 
have a geometric distribution defined by~(\ref{eq:probsplit}); moreover, we upper bound the number
of function evaluations using the number of elements on which the splitting is computed:
\begin{multline*}
\mathbf E\Biggl[\sum_{i=0}^{h-1} \sum_{j=0}^{s^i-1} X^i_j s^{h - i}\ell\Biggr]
= s^h\ell\sum_{i=0}^{h-1} \mathbf E\bigl[ X^i_0 \bigr] \\\sim
  s^h\ell\sum_{i=0}^{h-1} \sqrt{\frac{(2\pi)^{s-1} \bigl(s^{h - i - 1}\ell\bigr)^s }{s^{h-i}\ell}}
  \\= s^h\ell\sqrt{(2\pi \ell)^{s-1}} \sum_{i=0}^{h-1} \sqrt{\frac{\bigl(s^{h-i-1}\bigr)^s}{s^{h-i}}}
 \\\leq s^h\ell\sqrt{(2\pi \ell)^{s-1}} \sum_{i=0}^{h-1} \Bigl(\sqrt{s^{s-1}}\Bigr)^i
 \\=s^h\ell\sqrt{(2\pi \ell)^{s-1}} \frac{1 - \Bigl(\sqrt{s^{s-1}}\Bigr)^h}{1 - \sqrt{s^{s-1}}}
 \\\leq \sqrt{(2\pi)^{s-1}\bigl(s^h\ell\bigr)^{s+1}}.
\end{multline*}

As we already remarked, since the expected work on a bucket is polynomial in the bucket size,
the expected overall work for finding splittings will be linear in the number of keys. Our computation
does not apply to a generic splitting strategy, but the strategies we will use in practice
will be covered by our analysis.

\subsection{Searching for bijections}

Analogously, we can estimate
the probability of finding a MPHF on a leaf of size $m$. The
probability of hitting a bijection is $m!/m^m$, as there are $m!$ bijections
and $m^m$ overall possible functions from the leaf to itself,
so the average number of
trials is $m^m/m!\approx e^m/\sqrt{2\pi m}$ using Stirling's approximation.
%  Each
% trial will evaluate on average $\approx\sqrt{\pi m/2}$ times the hash
% function before finding a conflict~\cite{RamQ295,FGKRQF}, so we will assume
% that the number of evaluations per element in the leaf will be
% $\approx e^m/2m$.\footnote{We are here multiplying two expected values, so we
% cannot assume $e^m/2m$ to be anything more than a useful approximation.
% However, empirically it is within $20\%$ of the experimental value, which is
% more than sufficient for any practical purpose.}
Note that since leaf sizes
are bounded by a constant, the overall work for finding bijections is linear in the number of keys.

\subsection{Invariance}

A useful feature of our technique is that it satisfies an \emph{invariance} property:
the probability of finding a bijection on a set of $m$ elements is equal
to the probability of finding a splitting and finding a bijection on each part. More formally,

\begin{theorem}
\label{th:inv}
Given a set of $m$ elements, the probability of finding a minimal perfect hash on the set
is equal to the probability of finding a split in $s$ parts \lst ks, multiplied by the probabilities
of finding a bijection on each part.
\end{theorem}
\noindent\textit{Proof.}
Immediate, as
\[
\frac{m!}{m^m}\prod_{i=0}^{s-1}\frac{\bigl(k_i\bigr)^{k_i}}{k_i!} \prod_{i=0}^{s-1}\frac{k_i!}{\bigl(k_i\bigr)^{k_i}} = \frac{m!}{m^m}.\quad\Box
\]

As an inductive consequence, for \emph{every} tree defining splittings and bijections on a set of size $m$, the
product of the probabilities on all nodes is constant, and equal to $m!/m^m$.

% \begin{multline*}
% \mathbf E\Biggl[\sum_{i=0}^{h-1} \sum_{j=0}^{s^i-1} L^i_j \Biggr]
% = \sum_{i=0}^{h-1} s^i\mathbf E\bigl[ L^i_0 \bigr] \leq
%  \sum_{i=0}^{h-1} (s-1)\lg(2\pi)+ s\lg \bigl(s^{h-i-1}\ell\bigr)- \lg \bigl(s^{h-i}\ell\bigr)  - 1
% \\=  h(s-1)\lg(2\pi \ell) - h +\sum_{i=0}^{h-1}  s(h-i-1)\lg s - (h-i)\lg s
% \end{multline*}

\subsection{A splitting strategy}
\label{sec:strategy}

We start from the observation that if we had no limit on the construction time, searching
for a bijection would provide a space-optimal structure in expectation. This happens because
the optimal parameter $r(p)$ (i.e., the length of the fixed part) of a Golomb-Rice code~\cite{SolDC}
for a geometrically distributed source
with parameter $p$, i.e., $k\geq0$ appears with probability $(1-p)^kp$, is given by~\cite{KieSGPRC}
\begin{equation}
\label{eq:gr}
r(p) = \max\left\{\,0,\ceil*{\lg\left(-\frac{\lg\phi}{\lg(1-p)}\right)}\,\right\},
\end{equation}
where $\phi=\bigl(\sqrt5 + 1\bigr)/2$ is the golden ratio. For this choice, the expected length of the unary
part is
\[
\frac1{1-(1-p)^{2^{r(p)}}},
\]
which is always between $\phi$ and $1 + \phi$. In particular, in expectation the length of the codeword for
trials with probability $p$ when $p\to 0$ is
\begin{multline*}
1+\phi+ \ceil*{\lg\left(-\frac{\lg\phi}{\lg(1-p)}\right)} =
1+\phi+ \ceil*{\lg\frac{\ln\phi}{p+O\bigl(p^2\bigr)}} \\=
1+\phi+ \ceil*{\lg\ln\phi -\lg p - \lg(1+ O(p))} \\= c +\lg\frac1p + O(p),
\end{multline*}
with $1.56 \approx1+\phi+\lg\ln\phi \leq c < 2+\phi+\lg\ln\phi \approx 2.56$.
Essentially, modulo less than three bits (which we pay for not using full Golomb codes, for rounding and for instantaneousness)
the expected length of the codeword is the logarithm of its expected value. If we apply this formula to the case of
a bijection on $m$ elements, we obtain
\[
c +\lg\frac{m^m}{m!} + O\biggl(\frac{m!}{m^m}\biggr) = m\lg e + O(\lg m)
\]
so for large enough $m$ we can get arbitrarily close to the lower bound.\footnote{For
feasible leaf sizes, the per key expected cost of the unary prefix for bijections is $\gtrapprox 0.1$.}
 More interestingly,
the invariance property tells us that if we have probabilities \lst pt at the nodes of a splitting tree
for $m$ keys, then $\prod_{i=0}^{t-1}p_i=m!/m^m$. So now the cost of storing the tree is in expectation
\begin{multline*}
\sum_{i=0}^{t-1} \biggl(c + \lg\frac1{p_i} + O\bigl(p_i\bigr)\biggr) = ct + \lg\prod_{i=0}^{t-1}\frac1{p_i} + O\Bigl(\sum_{i=0}^{t-1}p_i\Bigr) \\= ct + m \lg e + O(\lg m).
\end{multline*}
In other words, we lose about $c$ bits for each split and bijection, but otherwise the number of bits will tend
to the optimal value when $m$ grows (and $t$ is fixed, which in particular means that
the leaf size has to grow). Increasing the number of splits will thus speed up
construction and slow down lookups, but just slightly increase the space usage, as long
as the number of keys going through the split is sufficiently large. The main space loss associated
with $c$, however, is that due to the leaves, as it is amortized over the smallest
number of keys.

Armed with the information gathered so far, we are now going to describe our
splitting strategy.
Our main drive is that of making the choice of the leaf size the main factor in
bounding the time required to build the data structure, and we will assume $\ell\leq 24$ as the search
on larger leaves is simply too slow; besides, the Golomb-Rice parameter in this range is smaller than $32$ and can
be packed into just five bits. We expect that larger
leaves will lead to less evaluations and thus to structures with faster lookups, too.

From~(\ref{eq:avgsplit}) it is clear that if we want to flatten the splitting tree,
we should use larger values of $s$ in the lower levels. We thus define the following criterion:
starting from the bottom, we want to aggregate leaves so that the work that is necessary to compute
the splitting is the same work as computing the leaf bijections, combined. We assume that each
trial for a splitting will need $m$ function evaluations, and that each trial for a bijection will require
$\sqrt{\pi m/2}$ function evaluations~\cite{RamQ295,FGKRQF}; both estimations are approximations.
If we look for the integer minimizing
\begin{equation}
\label{eq:firstaggr}
\biggl|s\sqrt{\frac{\pi\ell}2} \frac{\ell^\ell}{\ell!} - s\ell \frac{(s\ell)^{s\ell}}{(s\ell)!} \biggl(\frac{\ell!}{\ell^\ell}\biggr)^s\biggr|
\end{equation}
for $\ell\leq 24$ we obtain $s =\max\{\,2,\lceil 0.35\ell + 0.5\rceil\,\}$. Thus, we will greedily
try to aggregate this number of leaves.

If we go up another level, we can ask the same: that is, when the fanout is such that the work done is the same
as the work done on the two lower levels. This leads to minimizing
\[
\biggl|st\sqrt{\frac{\pi\ell}2} \frac{\ell^\ell}{\ell!}  - ts\ell \frac{(ts\ell)^{ts\ell}}{(ts\ell)!} \biggl(\frac{(s\ell)!}{(s\ell)^\ell}\biggr)^t\biggr|
\]
assuming $s$ is the (exact) solution of~(\ref{eq:firstaggr}). Once again, we can numerically compute the best integer solutions,
obtaining $t =\lceil 0.21\ell + 0.9\rceil$ for $7\leq \ell \leq 24$, and $2$ for $\ell < 7$.

One can continue with further aggregation levels using the same criterion, but the impact on the space used by the data structure
becomes negligible, and each aggregation level requires a test in the lookup code. After the second aggregation level we thus
fix the fanout at $2$, and use as unit (and left part) $\lceil\lfloor m/2\rfloor /st\ell\rceil \cdot st\ell$, with $s$ and $t$ as above.

\section{Data representation}

The following data must be stored to be able to evaluate a RecSplit MPHF efficiently:
\begin{itemize}
  \item The initial bucket-assignment function $g$.
  \item For each bucket, a representation of its splitting tree,
  including the indices stored at each node.
  \item Since we build MPHFs independently for each bucket, we need to store
  the \emph{prefix sums} of the number of keys in each bucket (i.e., for each bucket, the number
  of keys in all previous buckets).
  \item Finally, we will concatenate the representations of all buckets in a single bit array:
  we will thus need to store the \emph{offset} (i.e., the starting position) of each bucket in the array.
\end{itemize}

\subsection{The bucket-assignment function $g$}

First, we replace every key $x\in S$ with a unique signature $s(x)$ of $\Theta(n)$ bits (in our implementation, $128$).
The bucket assignment is generated using \emph{fixed-point
multiplication}: we interpret the value $u(x)$ of the upper $t$ bits (e.g.,
$t=64$) of $s(x)$ as a real number $\alpha(x)=u(x)/2^t$ in the interval
$[0\..1)$ represented in fixed-point arithmetic, and we assign to $x$ the bucket
$\bigl\lfloor \alpha(x) \cdot\lceil n/b \rceil\bigl\rfloor$.
If we interpret $\alpha(x)$ as a real random uniform value in the unit interval,
this operation correspond to an \emph{inversion}~\cite{DevNURVG} returning a
random uniform discrete value in $\lceil n/b \rceil$, which is exactly what
we need.\footnote{Indeed, fixed-point inversion 
has been since the early days the technique of choice for turning the bits returned by a pseudorandom number generator
into a uniform discrete value in a finite range (see, e.g.,~\cite{KnuMB}).} Since we use fixed-point
arithmetic, this amounts to computing $\bigl\lfloor u(x)\cdot\lceil n/b
 \rceil/ 2^t\rfloor$, which can be performed with a multiplication
and a shift. We will use the same technique when searching for splittings
and bijections.

\subsection{Splitting trees}

Splitting trees will form the bulk of the space occupied by a RecSplit instance. It is thus essential
to devise a parsimonious representation that is at the same time quickly accessible.

To attain these goals, we will not store the shape of the tree (e.g., pointers to children).
Instead, we only write the indices associated with each node in preorder using an
optimal Golomb-Rice code, in a bit array.
We remind that once the splitting strategy has been fixed,
the Golomb-Rice parameter of each index is known, as it depends only
on the number of keys associated with the node.\footnote{Empirically, using optimal Golomb codes reduces the size of the structure by less than $1\%$. The
fact that Golomb-Rice codes have a fixed part of constant length will have an important part
in the implementation.}

Since the number of
children of every node is defined by the splitting strategy,
we do not need to store them explicitly. However, when navigating the
tree top-to-bottom, whenever we do not move to the first child
of a node, we need to recursively skip the subtrees associated to
the previous children.

In principle, to skip a subtree, we only need to know
the Golomb-Rice parameter associated with the root: we then skip
the first code, and recurse into each child. This strategy would be, however,
very slow.

\begin{figure*}[t]
\centering
\includegraphics[scale=1]{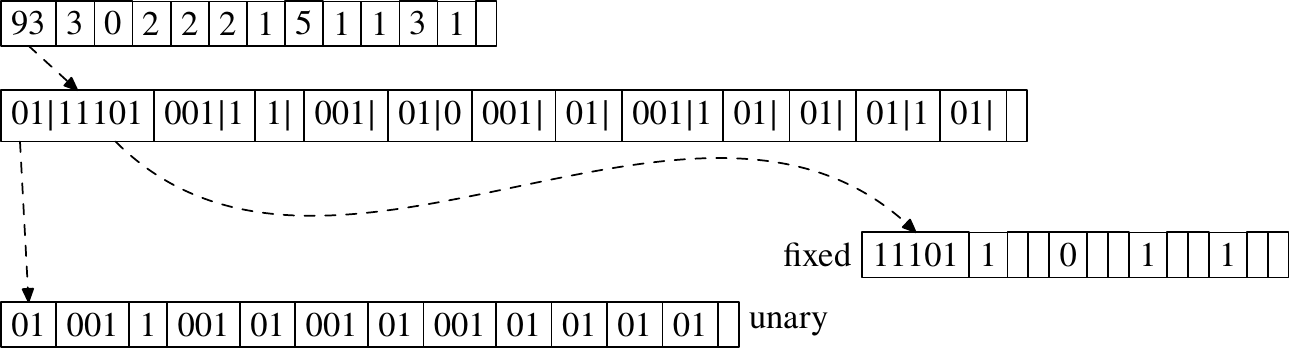}
\caption{\label{fig:codes}An example of coding for the tree in Figure~\ref{fig:tree}. The indices are first laid out
in preorder. Then each index is represented using a suitable Golomb-Rice code (the unary part is separated by a bar). Then
the unary and fixed parts are stored separately.}
\end{figure*}

We thus write all the constant-length part of the codes,
followed by all their unary parts (see Figure~\ref{fig:codes}) in the bit array. This choice makes skipping the
constant-length part of a subtree very fast, as we just have to move a pointer:
the amount of movement can be precomputed in a table (for a fixed splitting strategy, it depends only on the
number of keys associated with the node).
Using the same idea, we can retrieve the overall length of the fixed part, that is, the starting point of the unary codes, without additional information.

Skipping a number of unary codes is actually a \emph{selection} operation (e.g.,
find the $k$-th one in a bit array), for which very fast \emph{broadword
programming} algorithms exist~\cite{VigBIRSQ,GoPOSDSMD}.\footnote{On the Intel cores after Haswell
the PDEP instruction makes it possible to perform selection in a word using just three instructions.} We need to know the number of
ones to skip, which corresponds to the number of skipped nodes: also this number
can be precomputed and stored in a table.

\subsection{Prefix sums and offsets}

Finally, the prefix sums of bucket sizes $s_i$ and the offset of each bucket $o_i$
are stored using a customized Elias-Fano representation, which is a succinct
representation of monotone sequences~\cite{EliESRCASF,FanNBRISM}.

First, we compact our data by exploiting the dependence between these two values:
we store $s_i$ and $o_i-\beta s_i$, where $\beta$ is the number of bits per key
that are necessary to store the splitting trees.
Then, as it is customary, for both sequences we compute the minimum delta $\delta$ between
successive elements and use it to rescale the sequences by subtracting $i\delta$ from the $i$-th element:
this operation reduces the range of the sequences and, correspondingly, the bits per element.
If the minimum delta between items of the modified list  $o_i-\beta s_i$ is negative,
the rescaling will \emph{enlarge} the elements of the list so that it is again monotone.

The Elias-Fano representation uses space proportional to the logarithm of the
average gap between two elements, which in our case is $b$ or $\alpha b$. Thus, the space used to
store prefix sums can be reduced arbitrarily by enlarging the target bucket size $b$ (at the price
of slower construction and lookup).

\section{Implementation details}

\subsection{Logarithms}

In several computations we need to estimate the closest integer to $\lg
x$, that is, $\floor*{\lg x+1/2}$. For this, we use the approximate formula
\begin{multline*}
\lambda(x + (x \mathop\gg 1)) = \floor*{\lg(x+\floor*{x/2})} \approx \floor*{\lg(x+x/2)} \\= \floor*{\lg x + \lg 3- 1} \approx \floor*{\lg x + 0.58},
\end{multline*}
which contains only integer operations.

\subsection{Choosing the parameter for Golomb-Rice codes}

The optimal parameter $r(p)$ of a Golomb-Rice code for a geometrically
distributed source with parameter $p$ is given by~(\ref{eq:gr}).
The possible values of $p$ depend only on $m$, as the splitting strategy
specifies univocally the number and sizes of the parts in which to perform the split, so we can
precompute the possible values of $r(p)$ and store the results in a table for the most likely bucket sizes.\footnote{A single table of $32$-bit integers indexed by
bucket size is sufficient to memorize the Golomb-Rice parameter of a node, the skipping
information of its subtree and the number of nodes in the subtree when $\ell\geq 4$ and $b\leq 2000$.}

In the (practically negligible) case of large buckets we developed the following
integer approximation for the optimal parameter when splitting $m$ elements in
$s$ parts \lst ks:
\[
 \left(((s-1)\cdot 5 \gg 1) + \sum_{i=0}^{s-1}\lambda\bigl(k_i + \bigl(k_i
 \mathop\gg 1\bigr)\bigr) - \lambda(m)\right) \gg 1.
\]
Analogously, we store the optimal Golomb-Rice codes for bijections in a table,
as they will be needed only for a few dozen leaf sizes.
\subsection{Avoiding correlation}

For our estimations of the number of trials to be applicable, every search for a split or
for a bijection must be independent. We will compute both splittings and bijections
on the $\Theta(n)$-bit signatures used for the bucket assignments, but discard
the upper $t$ bits, so we will be working, in each bucket, with a set of random signatures.
When recursively descending during the splitting procedure, we take care of never
reusing functions that have been already searched through. Thus, along a path
independence is guaranteed by the independence of the functions $\phi^k_i$,
whereas on different parts it is guaranteed by the fact that the involved keys are actually
a random set of signatures with an empty intersection.

\subsection{Customizing Elias--Fano}

We briefly recall the details of the Elias--Fano representation. We assume to have a
monotonically non-decreasing sequence of $n>0$ natural numbers
\[
0\leq x_0\leq x_1 \leq \cdots \leq x_{n-2} \leq x_{n-1}\leq u,
\]
where $u>0$ is any upper bound on the last value.
We will represent such a sequence in two bit arrays as follows:
\begin{itemize}
  \item the lower $\ell=\max\{\,0,\lfloor \lg (u/n)\rfloor\,\}$ bits of each
  $x_i$ are stored explicitly and contiguously in the \emph{lower-bits
  array};
  \item the upper bits are stored in the \emph{upper-bits array} as a
  sequence of unary-coded gaps.
\end{itemize}
One then puts a \emph{selection} data structure on the high bits, so that
the position $p$ of the $k$-th one can be found in constant time. At that
point, the upper bits of the $k$-element are just $p-k$, and the lower
bits can be retrieved directly.% (see~\cite{VigQSI} for a more detailed explanation).

First, we notice that since the lower bits have fixed width,
if you need to work, as in our case, with \emph{two} Elias--Fano representations
of the same length and you always need to access the elements with the same
position in the two lists, the lower bits can be interleaved, saving one cache miss.

Then, we notice that the gaps between elements of the list are extremely regular,
as they are bucket sizes, or the number of bits used to encode a bucket. Because
of this property, the selection data structure can be significantly simplified
to a two-level inventory: we record the position of the
ones of index multiple of $2^{14}$ in a $64$-bit integer, and then use an array of $16$-bit integers
to record the position of the intermediate ones whose index is a multiple of $2^q$
for some $q$ (in our code, $q=8$). For the remaining ones we perform a local linear
search using broadword programming, starting from the closest one in the inventory.

By interleaving this information from the two lists,
we often save a further cache miss.

\section{Experiments}

In this section, we present the results of our experiments, which were performed
on an Intel\textregistered{} Core\texttrademark{} i7-7770 CPU @3.60GHz (Kaby
Lake), with 64\,GiB of RAM, Linux 4.17.19 and Java 12. For the C code, we
used the GNU C compiler 8.1.1.

Time is measured by wall clock. For the
lookup timings we report, the relative standard deviation is below
$5$\%. Construction times (which include reading the input,
generating the data structure and serializing it)
have a bit more variability due to I/O. We fix the CPU clock to avoid variations
due to throttling from the Turbo Boost controller.

We compare RecSplit with the three current state-of-the-art MPHF implementations:
\begin{itemize}
  \item GOV~\cite{GOVFSCF}\footnote{\url{https://github.com/vigna/Sux4J}} is a structure based on random hypergraphs that provides
  maps at $2.24$ bits per key with fast access.
  \item CHD~\cite{BBDHDC}\footnote{\url{https://sourceforge.net/projects/cmph/}} has been discussed in Section~\ref{sec:CHD}. It is by far the slowest
  map we tested, but it is the only one that can reach about $2$ bits per key.
  We report data for $\lambda=5$ and $\lambda=6$. We have not been able to build
  maps with $\lambda=7$ beyond a few thousand keys.
  \item BBHash~\cite{LRCFSMPHMKS}\footnote{\url{https://github.com/rizkg/BBHash}} is an implementation of fingerprinting-based minimal
  perfect hashing (see Section~\ref{sec:finger}) that aims at being very fast
  in construction and lookup,
  but uses a large amount of space. We tested the version using the smallest
  amount of space ($\gamma=1$), which however needs more than $3$ bits per key to be stored.
\end{itemize}

In our experiment, we build and evaluate maps on 128-bit random keys: this way,
we significantly increase the resolution of our results, as we cut off the time that
is necessary to compute hashes of long keys. Since all implementations we consider
hash as a first step their keys into random short keys, our choice of keys has 
no impact on the behavior of the structures.

\begin{figure}
\setlength{\abovecaptionskip}{-15pt}
\includegraphics[scale=1]{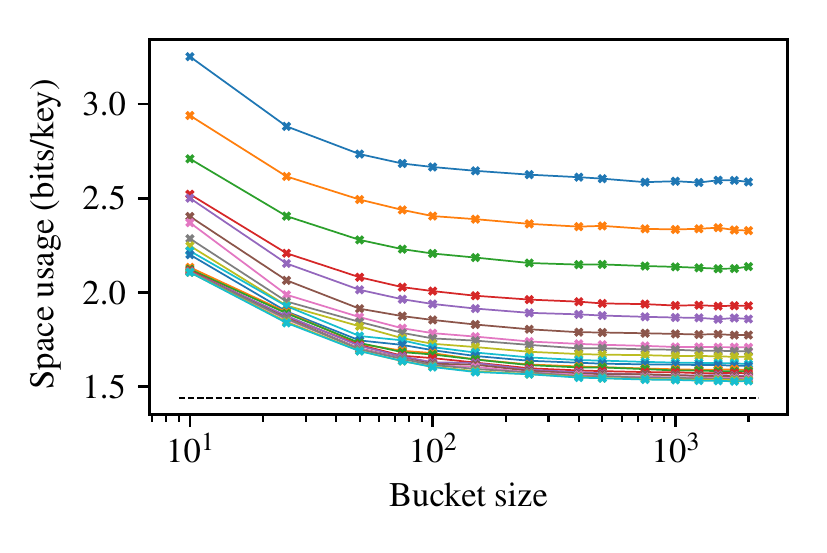}
\vspace*{1em}
\caption{\label{fig:space}Space usage depending on bucket size and leaf size. Lines from top
to bottom correspond to values of $\ell$ from $2$ to $21$. The dashed line shows the lower bound $\lg e\approx 1.44$.}
\end{figure}

RecSplit has different behavior depending on the parameters $\ell$ and $b$:
increasing $\ell$ leads to smaller data structures \emph{and}
faster lookups, at the price of a greater construction time. Increasing $b$ can
further decrease space (as the Elias--Fano lists are better amortized), at
the price of a slightly larger construction and lookup times. Figure~\ref{fig:space}
illustrates the dependence of space from these two parameters.

Among the many possible variations, we isolate:
\begin{itemize}
  \item $\ell=8$, $b=100$ breaks the $2$ bits/key barrier of CHD, providing
  better space, as well as significantly faster construction and lookup.
  \item $\ell=12$, $b=9$ uses less space than GOV and BBHash; it is faster
  or comparable in lookup time, but requires a longer time to build
  in a single threaded, non-distributed setting.
  \item $\ell=5$, $b=5$ uses less space than BBHash, and it is faster
  to build at large sizes.
  \item $\ell=16$, $b=2000$ is at the boundary of feasibility in construction, as
  it requires almost two milliseconds per key: nonetheless, it reaches $1.56$ bits per key,
  that is, $8.3\%$ from the lower bound. Its lookup time is significantly larger than
  that of GOV or BBHash, but it is still faster than CHD.
\end{itemize}

% TODO overlapping buckets
% would shave probably 0.1 b/key for the b<10
% we can try to fit it in for the final version

In Table~\ref{tab:1M} and~\ref{tab:1G} we report space usage in bits per key,
and the construction and lookup time in nanoseconds per key. In each group, we
compare an alternative (CHD, GOV, BBHash) with RecSplit. We remark that for all
structures we consider, the number of bits per key is essentially independent
from the size of the key set.

An immediate observation is that in the case of Table~\ref{tab:1M} all structures fit
the cache, whereas in the case of Table~\ref{tab:1G} they do not. As consequence,
the lookup times are an order of magnitude larger, even though all
structures perform lookups in constant time.

\begin{figure}
\setlength{\abovecaptionskip}{-15pt}
\includegraphics[scale=1]{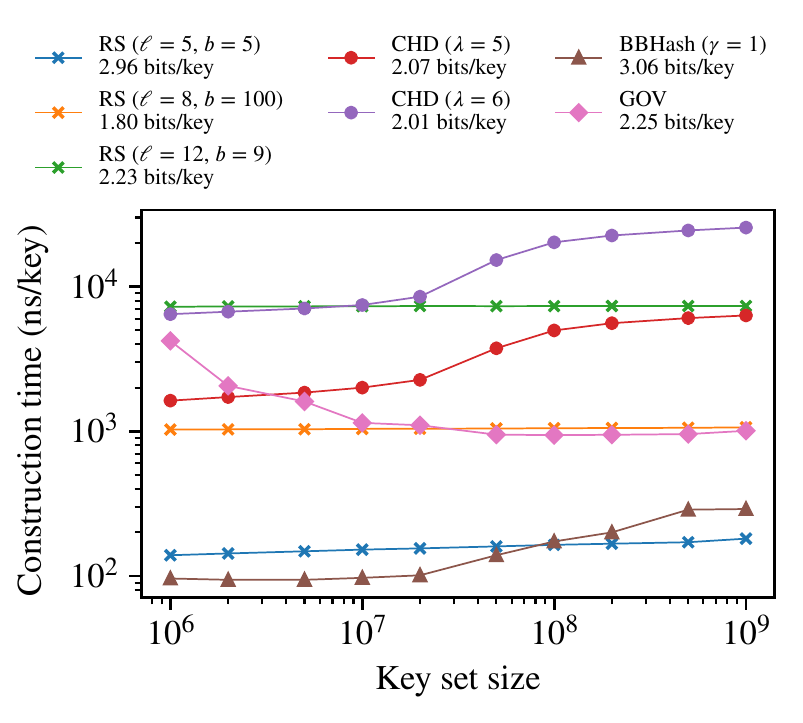}
\vspace*{1em}
\caption{\label{fig:construction}Construction time depending on the key set size.}
\end{figure}

\begin{figure}
\setlength{\abovecaptionskip}{-15pt}
\includegraphics[scale=1]{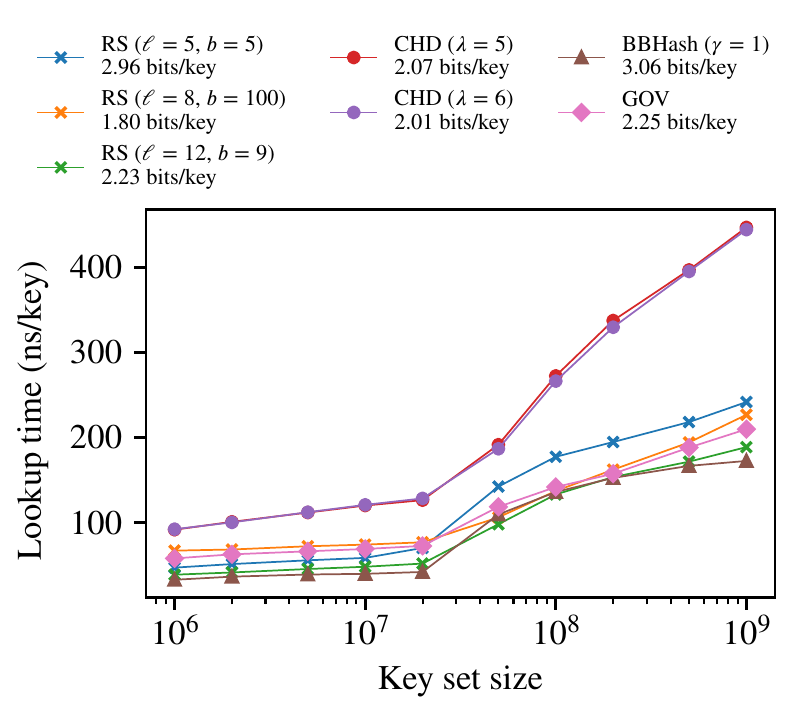}
\vspace*{1em}
\caption{\label{fig:lookup}Lookup time depending on the key set size.}
\end{figure}

A more global view
depending on the number of keys is given by Figure~\ref{fig:construction} and~\ref{fig:lookup}. Note that all code is in C or C++,
except for the construction of GOV, which is available only in Java.

\begin{table}
\renewcommand{\arraystretch}{1.4}
\begin{tabular}{lrrr}
MPHF&b/key&Constr.&Lookup\\\toprule
CHD ($\lambda=5$)	&	2.07	&	1627	&	91	\\
CHD ($\lambda=6$)	&	2.01	&	6440	&	92	\\
$\ell=8$, $b=100$	&	1.79	&	1027	&	67	\\\hline
GOV	&	2.26	&	4210	&	58	\\
$\ell=12$, $b=9$	&	2.23	&	7245	&	39	\\\hline
BBHash ($\gamma=1$)	&	3.07	&	96	&	33	\\
$\ell=5$, $b=5$	&	2.95	&	139	&	47	\\\hline
$\ell=16$, $b=2000$	&	1.56	&	1733801	&	87	\\
\end{tabular}
\caption{\label{tab:1M}Space usage, construction time and lookup time, in nanoseconds per key, for
a million keys.}
\end{table}

\begin{table}
\renewcommand{\arraystretch}{1.4}
\begin{tabular}{lrrr}
MPHF&b/key&Constr.&Lookup\\\toprule
CHD ($\lambda=5$)	&	2.07	&	6310	&	447	\\
CHD ($\lambda=6$)	&	2.01	&	25550	&	444	\\
$\ell=8$, $b=100$	&	1.80	&	1063	&	227	\\\hline
GOV	&	2.25	&	1007	&	210	\\
$\ell=12$, $b=9$	&	2.23	&	7331	&	189	\\\hline
BBHash ($\gamma=1$)	&	3.06	&	290	&	172	\\
$\ell=5$, $b=5$	&	2.95	&	181	&	241	\\
\end{tabular}
\caption{\label{tab:1G}Space usage, construction time and lookup time, in nanoseconds per key, for
a billion keys.}
\end{table}

\begin{figure}
\setlength{\abovecaptionskip}{-15pt}
\includegraphics[scale=1]{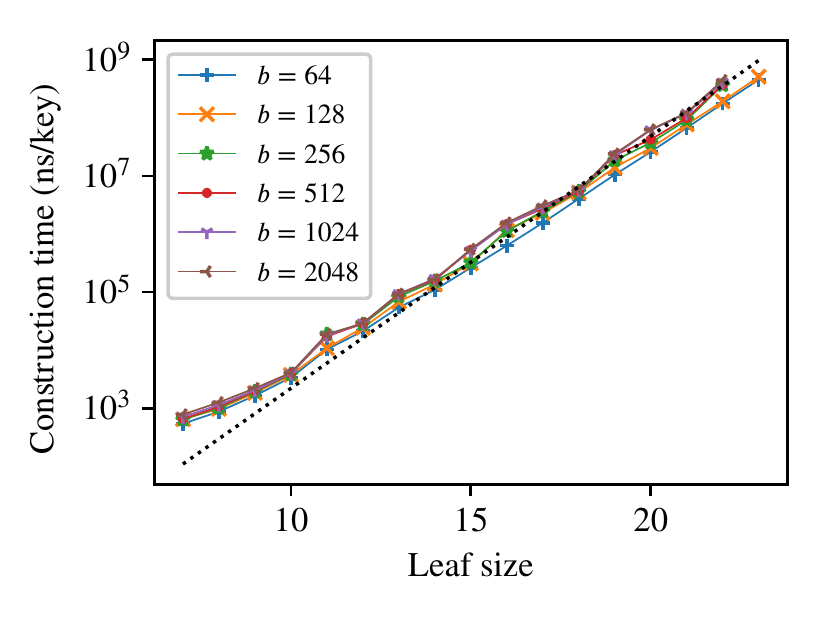}
\vspace*{1em}
\caption{\label{fig:constrleaf}Construction time per bucket size, depending on the leaf size. The dashed
line is $e^x/10$.}
\end{figure}

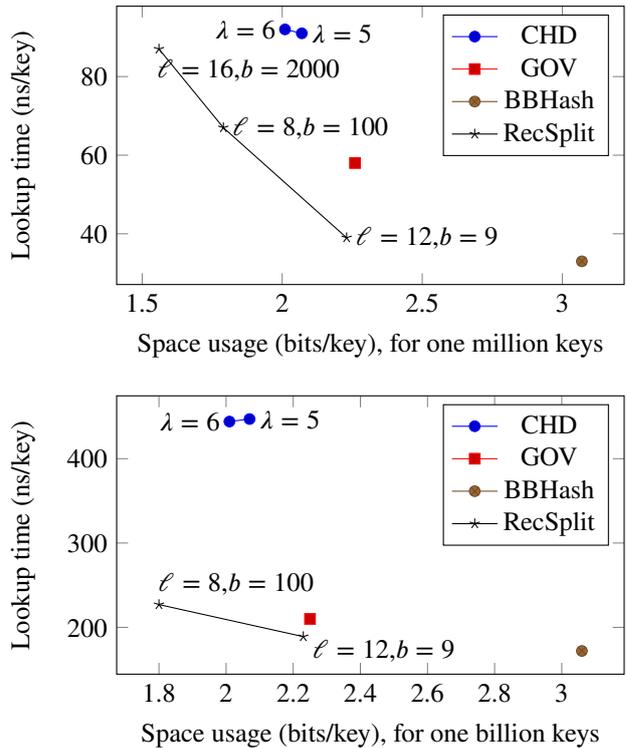
\begin{figure}
\setlength{\abovecaptionskip}{0pt}
\begin{tikzpicture}
\begin{axis}[
    width=235, height=150,
    legend pos=north east,
    mark=solid,
    mark size = 2pt,
    ylabel={Lookup time (ns/key)},
    xlabel={Space usage (bits/key), for one million keys}]
    \addplot plot coordinates {(2.07,91)(2.01,92)};
    \addlegendentry{CHD}
    \addplot plot coordinates {(2.26,58)};
    \addlegendentry{GOV}
    \addplot plot coordinates {(3.07,33)};
    \addlegendentry{BBHash}
    \addplot plot coordinates {(1.56,87)(1.79,67)(2.23,39)};
    \addlegendentry{RecSplit}
    \addplot[scatter, mark=none, only marks,
           point meta=\thisrow{color},
           nodes near coords*={\label},
               every node near coord/.style={anchor=\alignment},
           visualization depends on={value \thisrow{label} \as \label},
           visualization depends on=\thisrow{align} \as \alignment ]
table[meta=label] {
bits/key ns label color align
2.01 92 $\lambda=6$ 0 0
2.07 91 $\lambda=5$ 0 180
1.56 87 $\ell=16$,$b=2000$ 3 167
1.79 67 $\ell=8$,$b=100$ 3 180
2.23 39 $\ell=12$,$b=9$ 3 180
    };
\end{axis}
\end{tikzpicture}\vspace*{1em}
\begin{tikzpicture}
\begin{axis}[
    width=235, height=150,
    legend pos=north east,
    mark=solid,
    mark size = 2pt,
    ylabel={Lookup time (ns/key)},
    xlabel={Space usage (bits/key), for one billion keys}]
    \addplot plot coordinates {(2.07,447)(2.01,444)};
    \addlegendentry{CHD}
    \addplot plot coordinates {(2.25,210)};
    \addlegendentry{GOV}
    \addplot plot coordinates {(3.06,172)};
    \addlegendentry{BBHash}
    \addplot plot coordinates {(1.80,227)(2.23,189)};
    \addlegendentry{RecSplit}
    \addplot[scatter, mark=none, only marks,
           point meta=\thisrow{color},
           nodes near coords*={\label},
               every node near coord/.style={anchor=\alignment},
           visualization depends on={value \thisrow{label} \as \label},
           visualization depends on=\thisrow{align} \as \alignment ]
table[meta=label] {
bits/key ns label color align
2.01 444 $\lambda=6$ 0 0
2.07 447 $\lambda=5$ 0 180
1.80 227 $\ell=8$,$b=100$ 3 -165
2.23 189 $\ell=12$,$b=9$ 3 170
    };
\end{axis}
\end{tikzpicture}
\caption{\label{fig:pareto}Scatter plots of bits per key vs.~lookup time. Points on the Pareto
frontier are in the lower left region. The upper half is based on Table~\ref{tab:1M},
whereas the lower on Table~\ref{tab:1G}.}

\end{figure}

Immediately evident is that for each of the state-of-the-art
maps we compare with, \emph{there is an instantiation of RecSplit that is
comparable in space and/or lookup speed, and improves at least one of the two}.
Figure~\ref{fig:pareto} shows the \emph{Pareto frontier} (the set of
coordinate-wise minimal points) of the space of pairs giving space usage and
lookup time: below $3$ bits per element, these elements are all RecSplit instances. BBHash
has a slightly better lookup speed ($\approx 10\%$ faster) at the price of a $\approx 40\%$
larger space.

If we consider the three-dimensional Pareto frontier, the situation is different
because instances of RecSplit with fast lookup require usually more
construction time. However, RecSplit improves on all three parameters at the same
time with respect to CHD, which is presently the state of the art for small
space.
Moreover, both CHD and BBHash require that as much RAM as the
final size of the data structure is available. On the contrary, GOV and RecSplit
in principle allow to build the structure on-disk, possibly in a
distributed fashion, using a very small amount of RAM, as both structures
distribute keys in small buckets which can be processed independently.

Another interesting advantage of RecSplit is that, as we already mentioned, increasing $\ell$ leads to
slower construction, but at the same time decreases space and makes lookup faster, whereas
for BBHash, building a smaller map means slower construction and slower lookups;
for CHD, a smaller map means slower construction, and no effect on lookups.

If we consider the points associated with RecSplit instances with $\ell\geq 4$ and $64\leq b\leq 2048$ in isolation,
they are essentially all on the three-dimensional Pareto
frontier. This means that all these parameter choices provide different tradeoffs
between our three measures.

From the figures, it is evident that the in-memory construction of CHD and BBHash induces
a significantly nonlinear behavior due to cache and Translation Lookahead Buffer misses as
the number of keys increase. This does not happen for GOV and RecSplit.

Finally, in Figure~\ref{fig:constrleaf} we show the increase in construction time
depending on the leaf size for a choice of exponentially spaced bucket sizes. As
it is evident, increasing the leaf size by one approximately multiplies the construction
time by $e$, as expected from the discussion in Section~\ref{sec:strategy},
whereas the bucket size has less impact.

\section{Conclusions}

We have presented RecSplit, a new static data structure storing a minimal
perfect hash function with expected linear-time construction and expected
constant-time lookup. RecSplit is the first data structure able to break the
$2$\,bits/key barrier in practice, and instances very close to the lower
bound can be built feasibly, albeit slowly. Construction time can be reduced
by using a distributed computational setting such as
MapReduce~\cite{DeGMR}, as fixed-point inversion is monotone, so
buckets can be computed by a linear scan of the (offline) sorted signatures, and then
the splitting tree of each bucket can be built
independently.

By enlarging the size of the codomain of bijections and splittings, the
techniques described in this paper can also be used to build extremely compact
perfect (but not minimal) hash functions. We leave this extension for future
work.

\bibliography{biblio}
\end{document}